\documentclass[twocolumn,10pts]{article}
\newcommand{\br}{\mathbf{r}}
\newcommand{\be}{\mathbf{e}}
\newcommand{\bn}{\mathbf{n}}

\newcommand{\om}{\Omega}
\newcommand{\dom}{{\partial\Omega}}

\usepackage{bm,graphicx}

\title{\textbf{On the stress and torque tensors in fluid membranes}}
\author{Jean-baptiste Fournier
\\Laboratoire Mati\`ere et Syst\`emes Complexes (MSC),\\
Universit\'e Paris Diderot \& UMR 7057 CNRS\\
10, rue Alice Domon et L\'eonie Duquet, F-75205 Paris Cedex 13, France.\\
E-mail: jean-baptiste.fournier@paris7.jussieu.fr\\ \\
}
\date{Received XXXXth Month, 200X\\Accepted XXXXth Month,
200X\\DOI: 10.1039/}
\begin{document}

\maketitle
\renewcommand{\thefootnote}{\fnsymbol{footnote}}

\noindent \textbf{Abstract. -- We derive the membrane elastic stress and
torque tensors using the standard Helfrich model and a direct
variational method in which the edges of a membrane are infinitesimally
translated and rotated. We give simple expressions of the stress and
torque tensors both in the local tangent frame and in projection onto a
fixed frame. We recover and extend the results of Capovilla and Guven
[J.\ Phys.\ A, 2002, \textbf{35}, 6233], which were obtained using
covariant geometry and Noether's theorem: we show that the Gaussian
rigidity contributes to the torque tensor and we include the effect of a
surface potential in the stress tensor. Many interesting situations may
be investigated directly using force and torque balances instead of full
energy minimization. As examples, we consider the force exerted at the
end of a membrane tubule, membrane adhesion and domain contact
conditions.}

\bigskip

\section{Introduction}
\label{intro}

Lipid molecules dissolved in water spontaneously form bilayer membranes,
which may be idealized as fluid, incompressible surfaces with
superficial tension and bending rigidity~\cite{mouritsen_book}. These membranes,
which form vesicles or lamellar phases,  are widely studied in complex
fluid physics and in biology, as model systems of cell walls or
encapsulation agents, e.g., in the medical and cosmetic
industry~\cite{alberts_book}.

The elasticity associated with the deformations and fluctuations
of fluid membranes is given by a surface integral known as the Helfrich
Hamiltonian~\cite{helfrich73},
\begin{equation}\label{hel}
F=\int\!dA\left[\sigma+\frac{\kappa}{2}\left(c_1+c_2-c_0\right)^2
+\bar\kappa\,c_1c_2\right].
\end{equation}
The constant $\sigma$, usually called the ``surface tension", is better
understood as a chemical potential per surface unit fixing the value
of the (average) membrane area. Indeed, membranes have a fixed number
of lipids, hence essentially a fixed total area $A$. Instead of fixing
$A$, it is usually more convenient to let $A$ vary freely while adding a
term $\sigma A$ to the Hamiltonian in the canonical probability
distribution (very much like one adds a term $-\mu\,N$ in order to fix
the average value of $N$ in the grand-canonical ensemble). The last two
terms correspond to the most general quadratic curvature
energy for an isotropic surface. The parameters $c_1$ and $c_2$ are the
two principal curvatures, defined locally at each point of the surface
along two orthogonal directions.  The term with coefficient $\kappa$
(bending rigidity) favors a global curvature $c_1+c_2$ equal to some
constant $c_0$ characteristic of the membrane.  If the two monolayers
forming the membrane are identical, $c_0=0$ by symmetry, and the bending
term simply favors flatness. The term with coefficient $\bar\kappa$
(Gaussian rigidity constant) either favors spherical-like curvature,
i.e., $c_1 c_2>0$, if $\bar\kappa<0$, or saddle-like curvature, i.e.,
$c_1 c_2<0$, if $\bar\kappa>0$. Note that it is often omitted, since the
integral $\int dA\, c_1 c_2$ on a \textit{closed} surface only depends
on the topology of the surface (Gauss-Bonnet theorem).  

Using covariant differential geometry and Noether's theorem, Capovilla
and Guven~\cite{capovilla02} have recently derived the stress and torque
tensors associated with the Helfrich Hamiltonian. In this paper, we
revisit these two quantities. Although we use the so-called Monge gauge
for small deformations with respect to a flat membrane, our results in
the local tangent frame bear no difference with the general covariant
expression of~\cite{capovilla02}.  This is actually not surprising,
since the Helfrich Hamiltonian is quadratic in the local curvature. We
derive the stress and torque tensors using a direct and simple method
based on examining the elastic work done when translating or rotating
infinitesimally the edges of a membrane. We thus obtain the ``projected"
stress and torque tensors, all quantities referring to a basal plane
above which the membrane stands. We may then easily express all
quantities in the local tangent frame, thus obtaining very simple
expressions. Apart from the differences in the formulation, there are
three new results with respect to Ref.~\cite{capovilla02}: i) the
Gaussian rigidity ($\bar\kappa$) is shown to contribute to the torque
tensor, ii) the contribution of an external potential $V$ arising from a
substrate is discussed and included in the stress tensor, iii) the
stress and torque tensors are given both as ``local" and ``projected"
quantities, the latter formulation being useful when one needs to
integrate the force exerted by the membrane through an extended contour.

\section{Derivation of the stress and torque tensors}

The stress tensor, $\bm{\Sigma}$, relating linearly the force exerted by
the membrane through an infinitesimal cut to the vector normal to the
cut and proportional to its length is a tensor with $3\times2=6$
components~\cite{capovilla02}.  Indeed, the force is a three-dimensional
vector (usually not tangent to the surface), while the vector
normal to the cut may be taken as lying within the surface and thus
needs only be described by a two-dimensional vector. The same holds for
the torque tensor $\mathbf{T}$ which provides the elementary torque exchanged
through a cut.

\begin{figure}
\centerline{\includegraphics[width=.9\columnwidth]
{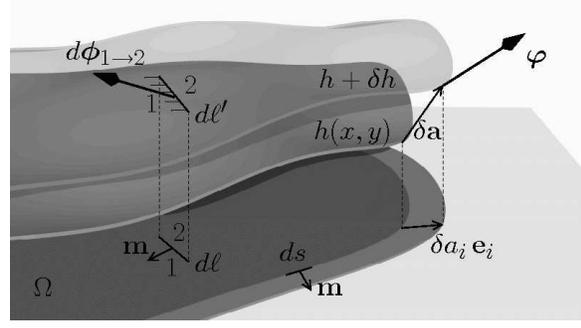}}
\caption{
Configuration and notations used in the derivation of the ``projected"
stress and torque tensors. The membrane $h(x,y)$ standing above domain
$\Omega$ (dark grey) is deformed and its edges are translated by
$\delta\mathbf{a}$. The deformed membrane, $h(x,y)+\delta h(x,y)$, is
drawn in transparency and its projection onto the basal plane is drawn
in a lighter grey. The force density $\bm{\varphi}$ acting along the
border and the force $d\bm{\phi}_{1\to2}$ exchanged through the cut
$d\ell'$ are shown (for the sake of clarity, the torques are not shown).
In the ``projected" formulation, the stress and torque tensors are
defined with respect to the \textit{projected} cut $\mathbf{m}\,d\ell$
lying on the basal plane instead of the actual cut $d\ell'$ lying
within the membrane.
}
\label{schema}
\end{figure}

Let us consider a membrane parametrized by its height $z=h(x,y)$ above a
reference plane $(x,y)$ in an orthonormal basis $(x,y,z)$.  Assuming
that the membrane is only weakly deformed with respect to the  
plane $(x,y)$, we shall neglect everywhere terms of order higher than
2 in the derivatives of $h$. Our aim, first, is to calculate the
``projected" stress tensor
$\bm{\Sigma}=\Sigma_{ij}\be_i\otimes\be_j+\Sigma_{zj}\be_z\otimes\be_j$
and the ``projected" torque tensor (per unit length)
$\mathbf{T}=T_{ij}\be_i\otimes\be_j+T_{zj}\be_z\otimes\be_j$ in the
fixed frame $(x,y,z)$.  Here and in the following, Latin indices stand
only for $x$ or for $y$.  These tensors are defined as follows. Consider
first an infinitesimal cut of length $d\ell'$ in the membrane,
separating two regions (see Fig.~\ref{schema}). Consider, next, the
\textit{projection} of this infinitesimal cut onto $(x,y)$, of length
$d\ell$ and normal $\mathbf{m}$ pointing towards the inside of
region~n$^\circ1$. By virtue of linearity, the infinitesimal force
$d\bm{\phi}_{1\to2}$ and the infinitesimal torque $d\bm{\tau}_{1\to2}$ that
region~n$^\circ1$ exerts onto region~n$^\circ2$ through the cut
$\mathbf{m}\,d\ell$ (for a
given membrane configuration) are given by
\begin{eqnarray}
d\bm{\phi}_\mathrm{1\to2}
\!\!&=&\!\!\mathbf{\Sigma}\cdot\mathbf{m}\,d\ell
\,\,=\,\, \mathbf{e}_i\,\Sigma_{ij}m_jd\ell
+\mathbf{e}_z\,\Sigma_{zj}m_jd\ell\,,\quad\\
d\bm{\tau}_\mathrm{1\to2}
\!\!&=&\!\!\mathbf{T}\cdot\mathbf{m}\,d\ell
\,\,=\,\, \mathbf{e}_i\,T_{ij}m_jd\ell
+\,\mathbf{e}_z\,T_{zj}m_jd\ell\,.
\end{eqnarray}
Summation over repeated indices will be implicit throughout. Note that
with the above sign convention, $\bm{\Sigma}$ can be considered as a
(tensorial) mechanical tension. 

\subsection{General derivation}

Consider now a deformed membrane (weakly departing from a plane) which
is at equilibrium under the action of external forces
$\bm{\varphi}$ and external torques $\bm{\tau}$ acting along its edges
(Fig.~\ref{schema}). Let $\Omega$ be the domain of $(x,y)$ above which
the membrane $h(x,y)$ is defined and $\partial\Omega$
its border, of curvilinear abscissa $s$ and outer normal $\mathbf{m}$.
The membrane elastic free-energy has the general form:
\begin{equation}\label{start}
F=\int_\om\!dx\,dy\,f(\{\partial_i h\},\{\partial_i\partial_k h\}
)\,.
\end{equation}

Calling $\br=(x,y)$, we vary the membrane shape: $h(\br)\to
h'(\br)=h(\br)+\delta h(\br)$, arbitrarily, while
translating the membrane edges by $\delta\mathbf{a}(\br)=\delta
a_i(\br)\,\mathbf{e}_i +\delta
a_z(\br)\,\mathbf{e}_z$ (recall that Latin indices stand only for $x$ or
$y$). We also apply to the
borderline normal
$\bn$ an infinitesimal rotation
$\delta\bm{\omega}(\br)=\delta\omega_i(\br)\,\mathbf{e}_i
+\delta\omega_z(\br)\,\mathbf{e}_z$. Integrating twice by parts, the
free-energy variation may be cast into the form :
\begin{eqnarray}
\label{dF}
\delta F\!\!&=&\!\!\int_\om\!dx\,dy\,
\frac{\delta{F}}{\delta h} \delta h+
\int_\dom\!ds\,m_i\Bigg[f\,\delta a_i+
\frac{\partial f}{\partial(\partial_i h)}\delta h
\nonumber\\
&&+
\frac{\partial f}{\partial(\partial_i\partial_k h)}\partial_k\delta h-
\partial_k\!\left(\frac{\partial f}{\partial(\partial_i\partial_k
h)}\right)\delta h
\Bigg]\,.
\end{eqnarray}
The border translation and rotation conditions imply
$\forall\br\in\dom$, $h'(\br+\delta
a_i(\br)\,\mathbf{e}_i)=h(\br)+\delta a_z(\br)$ and $\partial_k
h'(\br+\delta a_i(\br)\,\mathbf{e}_i)= \partial_k
h(\br)-\epsilon_{k\ell}\,\delta\omega_\ell(\br)
-\delta\omega_z(\br)\,\epsilon_{k\ell}\,\partial_\ell h(\br)$,
where $\epsilon_{k\ell}$ is the Levi-Civita
antisymmetric symbol. The latter relation follows from
$\delta\mathbf{n}=\delta\bm{\omega}\times\mathbf{n}$ and
$\mathbf{n}\simeq\mathbf{e}_z-(\partial_ih)\mathbf{e}_i$.
Hence, to first order in $\delta h$, we obtain the consistency relations
at the border :
\begin{eqnarray}
\label{r1}
\delta h\!\!&=&\!\!\delta a_z-\delta a_j\,\partial_j h\\
\label{r2}
\partial_k\delta h\!\!&=
\!\!&-\epsilon_{k\ell}\,\delta\omega_\ell
-\delta\omega_z\,\epsilon_{k\ell}\,\partial_\ell h-\delta
a_j\,\partial_j\partial_kh\,.
\end{eqnarray}

At equilibrium, the free-energy variation $\delta F$ must be
equal to the external work $\delta W$, which implies
\begin{equation}
\label{work}
0=\delta F-\delta W
=\delta F-\int_\dom\,ds\,(\bm{\varphi}\cdot\delta\mathbf{a}+
\bm{\tau}\cdot\delta\bm{\omega})\,.
\end{equation}
Using Eqs.~(\ref{dF})--(\ref{work}) and identifying the terms in factor
of $\delta a_j$, $\delta a_z$, $\delta\omega_\ell$ et $\delta\omega_z$,
which must vanish everywhere because of the arbitrariness of the shape
variation, we obtain 
$\varphi_j=m_i\,\Sigma_{ji}$, $\varphi_z=m_i\,\Sigma_{zi}$,
$\tau_\ell=m_i\,T_{\ell i}$ et $\tau_z=m_i\,T_{zi}$, yielding
\begin{eqnarray}
\label{Sigmaij}
\Sigma_{ij}&=&f\,\delta_{ij}
-\frac{\partial f}{\partial(\partial_j h)}\,\partial_i h
-\frac{\partial f}{\partial(\partial_j\partial_k
h)}\,\partial_i\partial_k h\nonumber\\
&+&\partial_k\!\left(
\frac{\partial f}{\partial(\partial_j\partial_k h)}
\right)\partial_i h\,,\\\label{Sigmazj}
\Sigma_{zj}&=&\frac{\partial f}{\partial(\partial_j h)}
-\partial_k\frac{\partial f}{\partial(\partial_j\partial_k h)}\,,\\
T_{ij}&=&\epsilon_{ik}\,
\frac{\partial f}{\partial(\partial_j\partial_k h)}\,,\\
\label{end}
T_{zj}&=&-\epsilon_{k\ell}\,
\frac{\partial f}{\partial(\partial_j\partial_k h)}\,\partial_\ell h\,,
\end{eqnarray}
which constitute the formal expressions of the ``projected" stress and torque
tensors.

\subsubsection{Stress tensor divergence}

Directly differentiating the components of the stress tensor yields
$\partial_j\Sigma_{zj}=-\delta F/\delta h$ and
$\partial_j\Sigma_{ij}=(\partial_ih)\,\delta F/\delta h$, where $\delta
F/\delta h=-\partial_i[\partial f/\partial(\partial_ih)]
+\partial_i\partial_j[\partial f/\partial(\partial_i\partial_jh)]$ is
the standard Euler-Lagrange term. We recognize in fact the
equation $\bm{\nabla}\cdot\bm{\Sigma}=-(\delta F/\delta h)\,\mathbf{n}$,
since at lowest order, the membrane normal is given by
$\mathbf{n}\simeq\be_z-(\partial_i h)\be_i$. This equation correctly
states that the restoring elastic force density exerted by the membrane is
$-(\delta F/\delta h)\,\mathbf{n}$. This is indeed a well-known starting
point in dynamical descriptions. At equilibrium, since $\delta F/\delta
h=0$, we obtain 
\begin{equation}
\partial_j\Sigma_{zj}=
\partial_j\Sigma_{ij}=0\,,
\end{equation}
i.e., the stress tensor is divergence-free.

\subsection{Case of the Helfrich Hamiltonian}

In the particular case of the Helfrich Hamiltonian~(\ref{hel}), we
obtain to second order in the derivatives
of $h$, the following elastic energy density: 
\begin{equation}
\label{helfrich}
f=\sigma+\frac{\sigma}{2}\left(\nabla h\right)^2
+\frac{\kappa}{2}\left(\nabla^2h-c_0\right)^2+\bar f\,,
\end{equation}
where $\bar
f=\bar\kappa\,\rm{det}(\partial_i\partial_jh)=\frac{1}{2}\bar\kappa\,[(\nabla^2
h)^2- (\partial_i\partial_jh)(\partial_i\partial_jh)]$ is the Gaussian
curvature contribution.
This yields $\partial
f/\partial(\partial_j h)=\sigma\,\partial_jh$ and $\partial
f/\partial(\partial_j\partial_kh)=(\kappa+\bar\kappa)\,\delta_{jk}\nabla^2h
-\bar\kappa\,\partial_j\partial_kh-\kappa\,c_0\,\delta_{jk}$. Since
$(\partial_i\partial_kh)\times\partial\bar
f/\partial(\partial_j\partial_kh)=-\bar f\delta_{ij}$ and
$\partial_k[\partial\bar
f/\partial(\partial_j\partial_kh)]=0$, we obtain explicitely
\begin{eqnarray}
\label{n1}
\Sigma_{ij}&=&
(f-\bar f)\,\delta_{ij}-\sigma
\left(\partial_i h\right)\left(\partial_j h\right)\nonumber\\
&-&\kappa\left(\nabla^2h-c_0\right)\partial_i\partial_j h
+\kappa\left(\partial_i h\right)\,\partial_j\nabla^2h\,,\\
\label{n2}
\Sigma_{zj}&=&\sigma\,\partial_j h-
\kappa\,\partial_j \nabla^2 h\,,\\
T_{ij}&=&\left(\kappa+\bar\kappa\right)\nabla^2h\,\epsilon_{ij}
-\bar\kappa\,\epsilon_{ik}\,\partial_j\partial_kh
-\kappa\,c_0\,\epsilon_{ij}\,,\quad\\
T_{zj}&=&-\left(\kappa+\bar\kappa\right)\left(\nabla^2h\right)
\epsilon_{j\ell}\,\partial_\ell h\nonumber\\\label{n4}
&+&\bar\kappa\,\epsilon_{k\ell}\left(\partial_j\partial_kh\right)\,
\partial_\ell h
+\kappa\,c_0\,\epsilon_{j\ell}\,\partial_\ell h\,.
\end{eqnarray}
Note that the contributions involving $\bar\kappa$ cancel altogether in
the expression of $\bm{\Sigma}$, but not in the expressions of
$\mathbf{T}$.

These four equations are our central result. Recall that they are valid
only up to second order in the derivatives of $h$. Recall also that they
give the components of the ``projected" stress and torque tensors : not
only the components are projected along the fixed basis $(x,y,z)$, but
also the elementary cut $m_j\,d\ell$ by which they must be multiplied
is, by definition, the projection onto the reference plane of a cut
within the membrane surface. 

More explicitly, the Cartesian components of the ``projected" stress
and torque tensors (in the case $c_0=0$ for the sake of simplicity) are
given by 
\begin{eqnarray}
\label{Sigmaxx}
\Sigma_{xx}&=&
\sigma
+\frac{\sigma}{2}
\left[\left(\partial_yh\right)^2-\left(\partial_xh\right)^2\right]
\nonumber\\
&+&\frac{\kappa}{2}
\left[\left(\partial^2_yh\right)^2-\left(\partial^2_xh\right)^2\right]
+\kappa\left(\partial_xh\right)\partial_x\nabla^2h\,,\quad\\
\label{Sigmaxy}
\Sigma_{xy}&=&
-\sigma\left(\partial_xh\right)\left(\partial_yh\right)
-\kappa\left(\partial_x\partial_yh\right)\nabla^2h
\nonumber\\
&+&\kappa\left(\partial_xh\right)\partial_y\nabla^2h\,,\\
\label{Sigmazx}
\Sigma_{zx}&=&\sigma\,\partial_x h-
\kappa\,\partial_x \nabla^2 h\,.
\end{eqnarray}
The other components follow by exchanging $x$ and $y$.
For the torque tensor, we obtain explicitly
\begin{eqnarray}
T_{xx}&=&-\bar\kappa\,\partial_x\partial_yh\,,\\
T_{xy}&=&\kappa\,\nabla^2h+\bar\kappa\,\partial^2_xh\,\\
T_{yx}&=&-\kappa\,\nabla^2h-\bar\kappa\,\partial^2_yh\,\\
T_{yy}&=&\bar\kappa\,\partial_x\partial_yh\,,\\
T_{zx}&=&-\kappa\left(\nabla^2h\right)\partial_yh\nonumber\\
&-&\bar\kappa\left(\partial_y^2h\right)\partial_yh
-\bar\kappa\left(\partial_x\partial_yh\right)\partial_xh\,\\
T_{zy}&=&\kappa\left(\nabla^2h\right)\partial_xh\nonumber\\
&+&\bar\kappa\left(\partial_x^2h\right)\partial_xh
+\bar\kappa\left(\partial_x\partial_yh\right)\partial_yh\,.
\end{eqnarray}

\subsection{Expressions in the tangent, principal frame}

Locally, for a membrane with a given fixed shape, it is always possible
to choose the frame $(x,y,z)=(X,Y,Z)$ in such a way that it is tangent
to the membrane and has $X$ and $Y$ oriented along the directions of
principal curvatures. For the sake of simplicity, we shall first
consider the case $c_0=0$. Hence, in the \textit{tangent, principal
frame}, we have
$\partial_Xh=\partial_Yh= \partial_X\partial_Yh=0$, and
$\partial^2_Xh=C_X$ and $\partial^2_Yh=C_Y$, where $C_X$ and $C_Y$ are
the principal curvatures.

\begin{figure}
\centerline{\includegraphics[width=\columnwidth]
{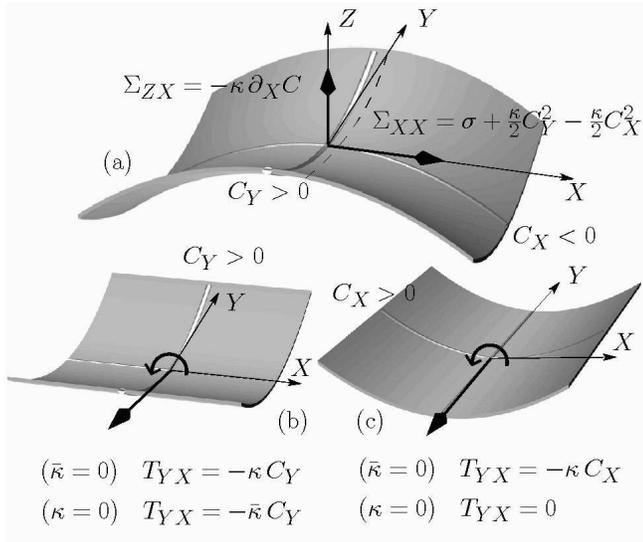}}
\caption{
(a) Components of the stress tensor in the tangent, principal frame (the
axes $X$ and $Y$ are oriented along the principal curvatures
directions), for $c_0=0$. The quantity $\Sigma_{XX}$ is the force per
unit length exerted along $X$ by the region $X>0$ onto the region $X<0$
through a cut normal to to the $X$ axis (i.e., parallel to the $Y$
axis).  Likewise, $\Sigma_{ZX}$ represent the component along $Z$ of the
same force. (b)--(c) In the tangent, principal frame, the torque
exchanged through a cut parallel to a principal direction is tangent to
the membrane and parallel to the cut. Several situations are depicted,
showing the contributions from $\kappa$ and $\bar\kappa$ (for
$\bar\kappa>0$ and $c_0=0$).
}
\label{parabo}
\end{figure}

For the stress tensor, either from Eqs.~(\ref{n1})--(\ref{n2}), or
Eqs.~(\ref{Sigmaxx})--(\ref{Sigmazx}), this yields
\begin{eqnarray}
\label{local}
\Sigma_{XX}&=&\sigma+\frac{\kappa}{2}C_Y^2-\frac{\kappa}{2}C_X^2\,,\\
\Sigma_{YX}&=&0\,,\\
\Sigma_{ZX}&=&-\kappa\,\partial_XC\,,
\end{eqnarray}
where $\partial_XC$ is the gradient along $X$ of the sum $C$ of the
principal curvatures. These expressions agree with the general covariant
formula of Ref.~\cite{capovilla02} (see also Ref.~\cite{muller05}). The
components $\Sigma_{YY}$, $\Sigma_{XY}$, $\Sigma_{ZY}$ simply follow by
exchanging $X$ and $Y$ in the above expressions. Hence, in the tangential
frame, the force exerted by the membrane through a cut parallel to a
direction of principal curvature has a tangential component
perpendicular to the cut and also a component normal to the membrane
(see Fig.~\ref{parabo}a).  In the case $c_0\ne0$, the contribution
$\Sigma_{XX}^{(0)}=\frac{1}{2}\kappa\,c_0^2-\kappa\,c_0\,C_Y$ must be
added to $\Sigma_{XX}$.

For the torque tensor, we get 
\begin{eqnarray}
T_{XX}&=&0\,,\\
\label{tyx}
T_{YX}&=&-\kappa\left(C_X+C_Y\right)
-\bar\kappa\,C_Y\,,\\
T_{ZX}&=&0\,.
\end{eqnarray}
The components $T_{YY}$, $T_{XY}$ and $T_{ZY}$ are obtained by
exchanging $X$ and $Y$ and multiplying by $-1$ (due to the presence of
the Levi-Civita antisymmetric symbol). In the case $c_0\ne0$, the contribution
$T_{YX}^{(0)}=\kappa\,c_0$ must be added to $T_{YX}$. 

The formula $\Sigma_{XX}=\sigma
+\frac{1}{2}\kappa\,C_Y^2-\frac{1}{2}\kappa\,C_X^2$ may be interpreted
as follows. It is given by the sum of the first and the third terms of
Eq.~(\ref{Sigmaij}) (the other two terms vanishing in the tangential
frame). The first term, disregarding the Gaussian contribution, is
$f\delta_{ij}=\sigma+\frac{\kappa}{2}(C_X+C_Y)^2$; the third term is
$-(\partial_i\partial_kh)\,\partial
f/\partial(\partial_j\partial_kh)=-\kappa(C_X+C_Y)C_X$. Indeed, if the
membrane is retracted in the direction perpendicular to the cut (dashed
line in Fig.~\ref{parabo}a), there is both an energy gain associated
with removing a band of membrane having an energy density $f$ (first term),
and, since this operation must be done at constant orientation of the
membrane normal to prevent torques work, a change in curvature energy if $C_X$
is non-zero (third term).

To check the validity of the formulas for the torque tensor, consider
first the case $\bar\kappa=0$ and $\kappa\ne0$, where
$T_{YX}=-\kappa(C_X+C_Y)$. As can be seen in Fig.~\ref{parabo}, the
effect of $T_{YX}$ (i.e., the tendency of the region $X>0$) is to curve
the cylinder into a saddle in Fig.~\ref{parabo}b or to reduce the
cylindrical curvature in Fig.~\ref{parabo}c : in both cases there is a
gain in bending energy since the mean curvature is lowered.  Next,
consider the case $\bar\kappa>0$ and $\kappa=0$, where
$T_{YX}=-\bar\kappa\,C_Y$. The tendency of the region $X>0$
is to curve the cylinder into a favorable saddle $(\bar\kappa>0)$ in
Fig.~\ref{parabo}b, while there is no torque in Fig.~\ref{parabo}c
because the Gaussian curvature is not affected by a rotation parallel to
the $Y$ axis. 

\subsection{Additional terms in the presence of an external surface
potential}

If the membrane is subject to an external potential arising from a
substrate, the Helfrich Hamiltonian~(\ref{hel}) is supplemented by an
``adhesion" term of the form $F_\mathrm{adh}=\int\!dA\,W(z)$ where $z$
is the membrane coordinate perpendicular to the substrate, and $W(z)$ a
short- or finite-range adhesion
potential~\cite{evans85,seifert90,seifert93}.  In the weak deformation
description where the membrane is parametrized by $h(\br)$, taking the
substrate itself as the reference plane, this gives $f\to
f+f_\mathrm{adh}$, with \begin{equation}\label{adh}
f_\mathrm{adh}=W(h)\left[1+\frac{1}{2}\left(\nabla h\right)^2\right]\,.
\end{equation} The free-energy density $f$ in~(\ref{start}) is now a
function of $h$ also, i.e., $f=f(h,\{\partial_i
h\},\{\partial_i\partial_k h\})$.  Repeating all the calculations
from~(\ref{start}) to~(\ref{end}) yields, however, exactly the same
formal results (\ref{Sigmaij})--(\ref{end}).

By direct differentiation of~(\ref{Sigmaij}) and~(\ref{Sigmazj}), we
obtain now $\partial_j\Sigma_{zj}=\partial f/\partial h-\delta F/\delta
h$ and $\partial_j\Sigma_{ij}=(\partial_ih)\,\delta F/\delta h$, with
$\delta F/\delta h=\partial f/\partial h-\partial_i[\partial
f/\partial(\partial_ih)] +\partial_i\partial_j[\partial
f/\partial(\partial_i\partial_jh)]$.  At equilibrium, since
$\delta F/\delta h=0$, it follows that
\begin{eqnarray}
\partial_j\Sigma_{zj}-\frac{\partial f}{\partial h}&=&0\,,\\
\partial_j\Sigma_{ij}&=&0\,.
\end{eqnarray}
The first equation is the balance of the forces along $z$
acting on a membrane element of projected area $dA_p=dx\,dy$. In this equation,
$\partial_j\Sigma_{zj}\,dA_p$ is the elastic part and $-(\partial
f/\partial h)\,dA_p= -W'(h)\,dA$ is the force exerted by the substrate on
the membrane
(note the inclination factor $dA/dA_p$). Since the force from the
substrate are only along $z$, there is no contribution in the second
equation.

Let us now examine how the explicit expressions~(\ref{n1})--(\ref{n4})
of $\bm{\Sigma}$ and $\mathbf{T}$ are modified when the Helfrich energy
is supplemented by the adhesion term~(\ref{adh}). Actually, setting $f\to
f+f_\mathrm{adh}$ is equivalent to replacing $\sigma$ by $\sigma+W(h)$
in $f$. Since in (\ref{Sigmaij})--(\ref{end}) no derivative is taken
with respect to $h$ itself, it follows that the
results~(\ref{n1})--(\ref{n4}) hold, provided $\sigma$ is replaced
everywhere by $\sigma+W(h)$. Hence, with
\begin{eqnarray}
f-\bar f&=&\left[\sigma+W(h)\right]\left[1+\frac{1}{2}\left(\nabla
h\right)^2\right]\nonumber\\
&+&\frac{\kappa}{2}\left(\nabla^2h-c_0\right)^2\,,
\end{eqnarray}
we obtain
\begin{eqnarray}
\Sigma_{ij}&=&
(f-\bar f)\,\delta_{ij}-\left[\sigma+W(h)\right]
\left(\partial_i h\right)\left(\partial_j h\right)\nonumber\\
&-&\kappa\left(\nabla^2h-c_0\right)\partial_i\partial_j h
+\kappa\left(\partial_i h\right)\,\partial_j\nabla^2h\,,\quad\\
\Sigma_{zj}&=&\left[\sigma+W(h)\right]\,\partial_j h-
\kappa\,\partial_j \nabla^2 h\,,
\end{eqnarray}
the expressions of the torque tensor being unchanged.

For instance, in the local tangent basis $(X,Y,Z)$, we find simply
\begin{eqnarray}
\label{local2}
\Sigma_{XX}&=&\sigma+W(h)+\frac{\kappa}{2}C_Y^2-\frac{\kappa}{2}C_X^2\,,\\
\Sigma_{YX}&=&0\,,\\
\Sigma_{ZX}&=&-\kappa\,\partial_XC\,,
\end{eqnarray}
i.e., the adhesion potential simply renormalizes the tension.

\section{Some useful applications}

\subsection{Force required to pull a tubule}

Membrane tubules can be spontaneously formed by pulling locally a
membrane~\cite{derenyi02,powers02}. From the Helfrich
Hamiltonian~(\ref{hel}), with $c_1=0$ and $c_2=1/r$ (and $c_0=0$), the
energy of a tubule with length $L$ and radius $r$ is equal to $F=2\pi r
L(\sigma+\frac{1}{2}\kappa/r^2)$.  Minimizing with respect to $r$ yields
the equilibrium radius $r^\star=\sqrt{\kappa/(2\sigma)}$. Then, the total
energy is $F=2\pi L\sqrt{2\kappa\sigma}$ and
the force required to pull the tube is thus
$\phi=dF/dL=2\pi\sqrt{2\kappa\sigma}$. It may be rewritten as
\begin{equation}
\label{factor2}
\phi=2\sigma\times2\pi r^\star\,.
\end{equation}
This factor of 2 is intriguing: naively, one would rather expect $\phi$
to be equal to the tension $\sigma$ multiplied by the contour length
$2\pi r$, curvature stress providing essentially normal forces. 

\begin{figure}
\centerline{\includegraphics[width=.8\columnwidth]
{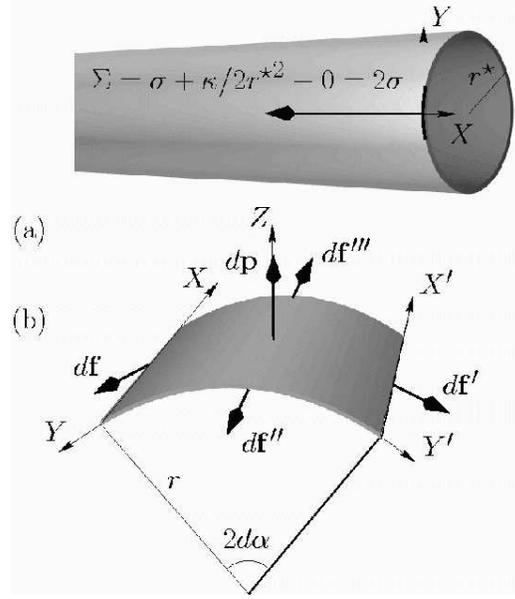}}
\caption{
(a) Elastic force $\Sigma$ per unit length exerted at the extremity of a
tubule with equilibrium radius $r^\star=\sqrt{\kappa/(2\sigma)}$. (b).
Scheme of the forces involved in the calculation of the equilibrium
radius of a tubule under pressure, using the Laplace law formalism with
the stress tensor.  }
\label{tube}
\end{figure}

Obviously, the stress tensor should give the answer. Consider a cut
along the membrane surface, perpendicular to the tube axis $X$, as
depicted in Fig.~\ref{tube}a. From Eq.~(\ref{local}), we obtain for the
tangential stress
\begin{equation}
\Sigma\equiv\Sigma_{XX}=\sigma+
\frac{\kappa}{2}\times\left(\frac{1}{r^\star}\right)^2
-\frac{\kappa}{2}\times0=2\sigma
\end{equation}
Note that the stress across a cut parallel to
the tube axis vanishes: $\Sigma_{YY}=\sigma-\frac{1}{2}\kappa/{r^\star}^2=0$.

It is interesting to see that the equilibrium tube radius can be deduced
from a generalized Laplace law. Assume that there is a pressure
difference $P$ across the membrane and consider a small patch of the
tube limited by four infinitesimal cuts along the principal curvature
directions (fig.~\ref{tube}b). The force along the $Z$ direction is
$dp=P\,2rd\alpha\,dX$.  Clearly, the forces $d\mathbf{f}''$ and
$d\mathbf{f}'''$ have no projection along $Z$. The contribution of the
forces $d\mathbf{f}$ and $d\mathbf{f}'$ along $Z$ gives
$P\,2rd\alpha\,dX=\Sigma_{YY}dX\,d\alpha+\Sigma_{Y'Y'}dX\,d\alpha$,
yielding
\begin{equation}
P=\frac{\frac{1}{2}\left(\Sigma_{YY}+\Sigma_{Y'Y'}\right)}{r}\,.
\end{equation}
Note that it is because the total curvature $C$ is
uniform that the normal components, e.g. $\Sigma_{ZX}$, do not
contribute.  Since
$\Sigma_{YY}=\Sigma_{Y'Y'}=\sigma-\frac{1}{2}\kappa/r^2$, we obtain the
equilibrium (Laplace) relation:
\begin{equation}
P=\frac{\sigma}{r}-\frac{\kappa}{2r^3}\,.
\end{equation}
This equation gives the correct equilibrium radius of the tube, as would
be obtained from minimizing $F=2\pi r
L(\sigma+\frac{1}{2}\kappa/r^2)-P\pi r^2L$ with respect to $r$. We also
obviously recover $r=r^\star$ when $P=0$.

\subsection{Adhesion and contact curvature}

\begin{figure}
\centerline{\includegraphics[width=.8\columnwidth]{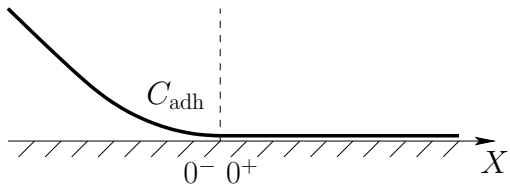}}
\caption{Membrane adhering onto a substrate. In the region $X>0$, the
membrane is flat and subject to the short-range potential
$W(h)=-W_\mathrm{adh}$.
In the limit where the adhesion potential is infinitely short-ranged,
the membrane departs tangentially in $X=0^-$ with a curvature
discontinuity $C_\mathrm{adh}$.
}
\label{adhesion}
\end{figure}

With the help of the stress tensor it is immediate to recover the
contact-curvature condition at the detachment point of an adhering
membrane~\cite{seifert90,capovilla02b}. Let us place ourselves in the
local tangent frame at the detachment point $X=0$~(Fig.~\ref{adhesion}).
For the sake of simplicity, we assume that the geometry is invariant in
the $Y$ direction, that the substrate is flat and that the adhesion
takes place very abruptly in $X=0$. The tangential component
$\Sigma_{XX}$ of the stress tensor is given by~(\ref{local2}). In
$X=0^+$, the membrane is flat ($C_X=C_Y=0$) and
$W(\left.h\right|_{0^+})=-W_\mathrm{adh}$, where $W_\mathrm{adh}$ is the
adhesion energy.
Hence,
\begin{equation}
\left.\Sigma_{XX}\right|_{0^+}=\sigma-W_\mathrm{adh}\,,
\end{equation}
In $X=0^-$, the membrane is still parallel to the substrate but it is
detached and
curved with the contact-curvature $C_X=C_\mathrm{adh}$. Assuming the limit of an
infinitely short-ranged potential, we may consider that
$W(\left.h\right|_{0^-})=0$. Hence,
\begin{equation}
\left.\Sigma_{XX}\right|_{0^-}=\sigma-\frac{\kappa}{2}C_\mathrm{adh}^2\,,
\end{equation}
The continuity of the tangential stress $\Sigma_{XX}$ yields then the
contact-curvature condition
\begin{equation}
C_\mathrm{adh}=\sqrt{\frac{2W_\mathrm{adh}}{\kappa}}\,.
\end{equation}

\subsection{Torque balance at the boundary between domains}

Vesicles made with different lipidic components may phase separate into
coexisting membrane domains~\cite{julicher96,baumgart03,allain04}. The
variational problem associated with the determination of the equilibrium
shape of a biphasic vesicule (even axisymmetric) is a difficult one,
requiring the introduction of Lagrange multiplier fields; however, it
yields a quite simple boundary conditions~\cite{julicher96}. Here we
show that this boundary condition may quite generally (and very easily)
be obtained from the continuity of the torque tensor.

\begin{figure}
\centerline{\includegraphics[width=.6\columnwidth]
{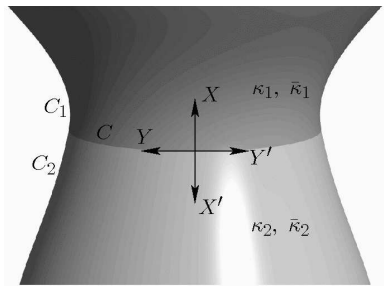}}
\caption{
Region near the phase separation line in an axisymmetric biphasic
vesicle. Geometry and notations used in the expression of the torque
balance through the
separation line.}
\label{vesicule}
\end{figure}

Consider an axisymmetric vesicle having a circular boundary
separating a phase with elastic constants $\{\kappa_1$, $\bar\kappa_1\}$
from a phase with elastic constants $\{\kappa_2$, $\bar\kappa_2\}$. We
place ourselves in the tangent frame $(X,Y)$ at a point along the
boundary, with $X$ normal to the boundary and pointing towards phase 1
(see Fig.~\ref{vesicule}). We define also the frame with opposite
directions $(X',Y')$. By symmetry, the axes $X$, $X'$, $Y$ and $Y'$ are
parallel to the principal curvature directions. Let $C_1$ (resp.\ $C_2$) be the
curvature of phase 1 (resp.\ 2) along $X$ (resp.\ $X'$)
and let $C$ be the common curvature of phases 1 and 2 along
either $Y$ or $Y'$. Then, from Eqs.~(\ref{tyx}), we obtain
\begin{eqnarray}
\frac{d\bm{\tau}^{1\to2}}{d\ell}&=&
T_{YX}^{(1)}\,\mathbf{e}_Y\,=\,
\left[
-\kappa_1\left(C_1+C\right)-\bar\kappa_1\,C
\right]\mathbf{e}_Y,\\
\frac{d\bm{\tau}^{2\to1}}{d\ell}&=&
T_{Y'X'}^{(1)}\,\mathbf{e}_{Y'}\,=\,
\left[
-\kappa_2\left(C_2+C\right)-\bar\kappa_2\,C
\right]\mathbf{e}_{Y'},\qquad
\end{eqnarray}
where $d\bm{\tau}^{1\to2}$ and $d\bm{\tau}^{1\to2}$ are the elementary
torques exchanged through
a cut of length $d\ell$ along the
boundary. At equilibrium these two torques must balance, i.e.
$d\bm{\tau}^{1\to2}+d\bm{\tau}^{2\to1}=0$. This yields
\begin{equation}
\kappa_2\,C_2-\kappa_1\,C_1=\left(\kappa_1+\bar\kappa_1-\kappa_2-\bar\kappa_2
\right)C\,,
\end{equation}
which is precisely the boundary condition (A22) established in the appendix
of Ref.~\cite{julicher96}.

\section{Summary}

The principal aim of this paper is to make it straightforward for
scientists working in the field of membranes to use stress and torque
concepts (instead of systematic energy minimization). To this end, from
the standard Helfrich elasticity, we have derived the stress and torque
tensors in a simple manner yielding explicit formulas. We have derived
the expressions of the stress and torque tensors both with respect to a
fixed ``projected" frame and in the local tangent frame.  Although we
have restricted our calculations to small deformations with respect to a
flat shape, our results are compatible with the fully covariant results
of Ref.~\cite{capovilla02}: this is because one may always work locally
in the tangent plane and because the Helfrich energy is only quadratic
in the curvature. We have included the contribution arising from the
Gaussian rigidity, which is always present and cortributes to the torque
tensor; we have included the contribution from a possible surface
adhesion potential.  We have shown several examples where the direct use
of stress and torque concepts is very fruitful.

\medskip
Stimulating discussions with P. Sens, at the origin of this work, are
gratefully acknowledged. The author also thanks J.-F. Joanny for
enlightening discussions.


\end{document}